\documentclass[useAMS, usenatbib, amssymb]{mnras}
\usepackage{psfig}
\usepackage{graphicx}
\usepackage{epsf}
\usepackage{bm}
\usepackage{lscape}
\usepackage[usenames, dvips]{color}
\usepackage[T1]{fontenc}
\title [On the Zero Point Constant of the BC Scale]
{On the Zero Point Constant of the Bolometric Correction Scale}
\author[Eker~et~al.]
       {Z. Eker $^{1}$\thanks{E-mail: eker@akdeniz.edu.tr},
V. Bak{\i}\c{s}$^{1}$, F. Soydugan$^{2,3}$ and S. Bilir$^{4}$
\\
    $^1$Akdeniz University, Faculty of Sciences, Department of Space Sciences and 
Technologies, 07058, Antalya, Turkey\\
  $^2$Department of Physics, Faculty of Arts and Sciences, \c{C}anakkale Onsekiz 
Mart University, 17100 \c{C}anakkale, Turkey\\
  $^3$Astrophysics Research Center and Ulup{\i}nar Observatory, \c{C}anakkale 
Onsekiz Mart University, 17100, \c{C}anakkale, Turkey\\
  $^4$Istanbul University, Faculty of Science, Department of Astronomy and Space
Sciences, 34119, Istanbul, Turkey\\
}
\date{}

\pagerange{\pageref{firstpage}--\pageref{lastpage}} \pubyear{}

\begin{document}

\maketitle

\label{firstpage} 

\begin{abstract}
Arbitrariness attributed to the zero point constant of the $V$ band bolometric corrections ($BC_V$) and its relation to ``bolometric magnitude of a star ought to be brighter than its visual magnitude'' and ``bolometric corrections must always be negative'' was investigated. The falsehood of the second assertion became noticeable to us after IAU 2015 General Assembly Resolution B2, where the zero point constant of bolometric magnitude scale was decided to have a definite value $C_{Bol}(W)= 71.197~425~...$~. Since the zero point constant of the $BC_V$ scale could be written as $C_2=C_{Bol}-C_V$, where $C_V$ is the zero point constant of the visual magnitudes in the basic definition $BC_V=M_{Bol}-M_V=m_{bol}-m_V$, and $C_{Bol}>C_V$, the zero point constant ($C_2$) of the $BC_V$ scale cannot be arbitrary anymore; rather, it must be a definite positive number obtained from the two definite positive numbers. The two conditions $C_2>0$ and $0<BC_V<C_2$ are also sufficient for $L_V<L$, a similar case to negative $BC_V$ numbers, which means that ``bolometric corrections are not always negative''. In sum it becomes apparent that the first assertion is misleading causing one to understand bolometric corrections must always be negative, which is not necessarily true. 
\end{abstract}


\section{Introduction}
One of the most very basic definitions in stellar astrophysics is bolometric correction ({\it BC}) known as the difference between the bolometric and visual magnitudes of a star ($BC_V = M_{Bol}-M_V=m_{Bol}-m_V$). Preference of the $V$ filter is simply a convention since most (including the oldest) photometric data is in the visual. 

After giving its basic definition as $BC_V=m_{Bol}-V$ and stating ``... this definition is usually interpreted to imply that the bolometric corrections must always be negative, although many of the currently used tables of empirical $BC_V$ values violate this condition'', \citet*{Torres10} paid attention to another definition of $BC_V$,

\begin{equation}
BC_V = 2.5\log \left(\frac{\int_0^\infty \!S_{\lambda}(V)f_{\lambda}\rm d\lambda}{\int_0^\infty \!f_{\lambda}\rm d\lambda}\right)+C_2,  
\end{equation}
where $S_{\lambda}(V)$ is the sensitivity function of the $V$ magnitude system, $f_{\lambda}$ is the monochromatic flux from a star, and $C_2$ is an assumed arbitrary constant of integration. Consequently, another problem was pointed out: ``The constant $C_2$ contains an arbitrary zero point that has been a common source of confusion''. This is because ``When making use of tabulations of empirical bolometric corrections for stars ($BC_V$), a commonly overlooked fact is that while the zero point is arbitrary, the bolometric magnitude of the Sun ($M_{Bol,\odot}$) that is used in combination with such tables cannot be chosen arbitrarily. It must be consistent with the zero point of $BC_V$ so that the apparent brightness of the Sun is reproduced'' as reported by \citet{Torres10}. 

Confirming \citet{Torres10}, \citet{Casagrande18} stated that despite the solar luminosity being a measured quantity, $M_{Bol,\odot}$ is an arbitrary zero point and any value is equally legitimate on the following equation

\begin{equation}
M_{Bol}-M_{Bol,\odot}=-2.5\log (L/L_{\odot}),
\end{equation}
on the condition that once chosen, all bolometric corrections are scaled accordingly. Therefore, equation (2) indicates the existence of an infinite number of $M_{Bol}$ and $M_{Bol,\odot}$ pairs having the same magnitude difference ($M_{Bol}-M_{Bol,\odot}$) to compute the single stellar luminosity ($L/L_{\odot}$) in solar units. Consequently, an infinite number of $BC_V$ is inevitable for a single star because there is no similar mechanism to change its $M_V$ parallel to its $M_{Bol}$ in order to keep $BC_V$ constant in the basic definition ($BC_V=M_{Bol}-M_V$). Therefore, according to \citet{Casagrande18}, an infinite number of zero points for the bolometric magnitudes and for {\it BC} scales must exist and all must be valid. 

Arbitrariness attributed to the zero point of the $BC_V$ scale was claimed long before by \citet{Hayes78} because all the $BC_V$ values tabulated by \citet{Hayes78} are less than zero and his statement ``I have taken the zero of B.C. to be at its minimum absolute value, with the result that $B.C._{\odot}=-0^{\rm m}.14$'' clearly indicates arbitrariness of the {\it BC} scale, where the smallest of the corrections, $BC_V=-0.01$, are associated with main sequence stars of the A7-F7 spectral types. Similarly, a tabulated empirical $BC_V$ table by \citet{Habets81} does not contain any positive value where $BC_V$ for the sun is -0.34 mag. The smallest correction, $BC_V=-0.07$, is associated with F2 spectral type. $BC_V$ by \citet{Kuiper38}, who preferred to take the zero point at 6600 K, are the same; all are less than zero. \citet{McDonald52} adjusted their corrections to a scale giving $BC_V=-0.11$ mag for the Sun with $T_{\odot}=5713$ K. \citet{Wildey63} evaluated all $BC_V$ determinations and presented them in a two-dimensional diagram $(B-V)_{\rm 0}-BC_V$, where all {\it BC} values appear less than zero. Another author, who gave $BC_V$ values less than zero, is \citet{Popper59}. The $BC_V$ values of pre-main-sequence stars are all given in the negative by \citet{Pecaut13}. The handbook of astronomers, Allen's Astrophysical Quantities, contains $BC_V$ numbers for main-sequence, giants, and supergiants which are all negative, where the smallest of the corrections, $BC_V=-0.00$ mag, is associated with F2 supergiants, and $T_{eff}=7030$ K \citep{Cox00}. Considering his note on page 381, where the definition of {\it BC is given} by \citet{Cox00}, the negative sign appears to be intentional to emphasize that all $BC_V$ must be negative.     

However, the picture is totally different among the other half of researchers, who use the same basic definition, and the alternative definition used by \citet{Torres10}. 

According to \citet{Heintze73}, $f_{\lambda}$ for the Sun is a measurable quantity over a sufficiently large wavelength region, thus $BC_{V, \odot}$ is known from the basic definition, $M_{Bol, \odot}$ and the zero point was adopted for the bolometric magnitude scale. So, $C_2$ is known \citep{Heintze73}. Confirming this, \cite{Code76} say that ``$C_2$ is uniquely defined if we adopt a value for the bolometric correction for a specified spectral distribution of $f_{\lambda}$''. Using not only the Sun's spectral distribution, but also the spectral distribution of 21 non-reddened single stars with well-established $\int_0^\infty \!S_{\lambda}(V)f_{\lambda}\rm d\lambda$, \citet{Code76} determined the value of $C_2=0.958$. \citet{Heintze73} did not give any table, but \citet{Code76} gave the tabulated $BC_V$ as being less than zero ($BC_V<0$) for main-sequence stars in the effective temperature range 5780-34000 K, except for a limited portion  in the 7000-8000 K range with $BC_V = 0.01$.

The table of \citet{Code76} is not the only example to include a limited number of positive $BC_V$ values. The tables presented by \citet{Johnson64, Johnson66}, \citet{Flower77, Flower96}, \citet*{Bessell98}, \citet{Sung13} and recently \citet{Casagrande18} and \citet{Eker20} all contain a limited number of positive $BC_V$ values.

Clearly, there are two schools of thought who treat arbitrariness of the $BC_V$ scale differently. One school intentionally shifts computed $BC_V$ values to make them all negative while the other school prefers retaining them as they are, even if there are positive $BC_V$ values. The first group appears self consistent according to the principle that a positive $BC_V$ would violate the fact that the bolometric magnitude of a star ought to be brighter than its $V$ magnitude; therefore, this group's members feel free to shift the zero point of the $BC_V$ scale to make sure that all $BC_V$ are negative. However, the second group, which retains the $BC_V$ as they are, even if there are positive values, either ignores the note ``$BC=m_{Bol}-V$ (always negative)'' \citep[page 381 in][]{Cox00} or a positive $BC_V$ is not seen as a violation. If the zero point of the {\it BC} scale is indeed arbitrary, why not take advantage and use it to make all {\it BC} values negative in order to avoid a conceptual problem?

Here we should stress the difference between absolute and limited discrete arbitrariness. Absolute arbitrariness is the arbitrariness expressed by \citet{Torres10} on equation (1) where $C_2$ could be any number from minus infinity to plus infinity, which would all be legitimate, and satisfy the solution. However, for the limited discrete arbitrariness, a researcher does not have the freedom to choose any value except a limited number of choices. The former is problematic while, the latter option, is acceptable. One step after limited discrete arbitrariness is the nominal choice (or standard choice). That is, standardization; if all users have a single option. Although the arbitrariness by \citet{Casagrande18} gives the impression of absolute arbitrariness, after reading why they choose $M_{Bol,\odot}= 4.74$ mag, it becomes clear why this is not an absolute but a limited discrete arbitrariness. 

Motivated to try to solve the dilemma between these two schools of thought, in this study we investigated the problem of arbitrariness on the $BC_V$ scale and its relationship to the statements ``the bolometric magnitude of a star ought to be brighter than its visual magnitude'' and ``bolometric corrections must always be negative''.  

\section{Groundwork}
\subsection{Non-Uniqueness (Arbitraryness) Anaysis}

One of the most distinguished philosophers of science in the 20th century, Sir Karl \citet{Popper72}, says that if a theory explains everything, it has no scientific value. Similarly, if a scientific problem has an infinite number of solutions (absolute arbitrariness),  such as solutions of an indefinite integral with an arbitrary constant, it would be equivalent to no solutions at all. Therefore, before discussing the problems associated with a positive {\it BC}, it is necessary to analyze the non-uniqueness (arbitrariness) attributed to the {\it BC} scale. Unfortunately, the word ``unique'' does not have a unique meaning. Thus, one must  first be aware of already identified types of non-uniqueness (arbitrariness) too. Similar analysis had been done for the synthetic light curves of spotted stars \citep{Eker99} and main-sequence mass-luminosity (MLR), mass-radius (MRR) and mass-effective temperature (MTR) curves \citep{Eker18}, where three types of non-uniqueness (Type 1, Type 2, Type 3) were investigated. 

Type I applies to a scientist who has already collected data (measurements, observations) and is at the point of trying to find a best-fitting curve (function), which could be picked from a list of many mathematical functions. For most cases, however, mathematical functions are known; e.g., a Planck function is for explaining the spectral energy distribution (SED) of a star. A problem with Type 1 exists if scientists do not have any notion about which function to apply and there could be more than one, perhaps many different mathematical functions that equally explains the data. 

If a Type 1 problem does not exist, the next problem is called Type 2, which concerns whether a unique fit is possible or not. Nevertheless, there are methods how to deal with Type 2 problems; e.g., the least squares method guarantees a unique fit for a given function; free parameters and coefficients can be determined even with their associated errors. In the case of SED, the free parameter is effective temperature of the star, which is a unique (non-arbitrary) value.

Lastly, the problem of Type 3, involving parameter space, needs to be investigated if there are no problems of Type 1 and Type 2. This is because there could be more than one set of parameters (discrete arbitrariness), even an infinite number of sets (unscientific case), that may produce the same curve. A good example of limited arbitrariness would be the light curve of a single circular spot on a rotating spherical star. Here, there is a trade off between the inclination of the rotation axis and the spot latitude; that is, a circular spot at latitudes 10 degrees (spot position with respect to the equator) on a spherical rotating star with 45 degrees of inclination produces exactly the same light curve as another spherical star with 80 degrees of inclination having the same size circular spot at 45 degrees latitudes \citep{Eker96}. A spot modeler cannot differentiate between these two models from the light curves unless the inclination of the rotation axis is pre-assumed or known. On the other hand, since each of the MLR, MRR and MTR functions \citep{Eker18} has one to one correspondence (one luminosity, one radius, one effective temperature for a main-sequence star of a given mass), there is no Type 3 problem for the interrelated MLR, MRR and MTR functions. 

When discussing these types of arbitrariness, \citet{Eker99} and \citet{Eker18} assumed that the scientific data (observed light curves, stellar masses and parameters respectively) did not involve any kinds of zero point complexities. The arbitrariness attributed to the zero point constant of the {\it BC} scale, if {\it BC} numbers are treated as measurements (data), is another kind of non-uniqueness, which is a totally new case in addition to the non-uniqueness (arbitrariness) problems discussed above. Since this new problem is directly connected with the measurements (data), it should be called Type 0 and should be tested or investigated before the other types. One may imagine the mass of stars being expressed by unknown units in such a way that it is impossible to convert them into a known unit, because the scales used by different observers have different zero points. The result is that the data would be unusable.

Science has overcome these problematic issues so that today's scientist enters a world where all units and unit systems (SI, cgs, etc) in science are ready and rules how to convert from one system to another already exist. In one way or another, the determination of a standard zero point constant of the {\it BC} measurements was completed only very recently. Another peculiarity is that the definition of {\it BC} in equation (1) involves definite integrals. The definite integrals, however, never take an arbitrary constant. Why such an arbitrary constant $C_2$ required in equation (1) is one of our basic questions, but let us leave this issue aside for the moment and start investigating whether there is a way out of the absolute arbitrariness implied by the arbitrary zero point constant $C_2$. It is clear to us since there is neither a boundary condition to apply nor could it be invented to set up its own value, that the only way out of the absolute arbitrariness is a consensus, as  was done historically for the fundamental units of length, mass and time. Here, it would be useful to review earlier attempts to avoid absolute arbitrariness before arriving at a final solution. 

\subsubsection{Early Attempts to Avoid Absolute Arbitrariness}

There are two possible ways of breaking off the absolute arbitrariness of the {\it BC} scale using a star, the Sun, despite the difficulty of measuring its apparent visual magnitude, which is about 26 mag brighter than the brightest star in the sky, and unlike stars the Sun is not a point source. 

1- Assume a bolometric correction ($BC_{V,\odot}$) for the Sun, and then calculate the solar bolometric magnitudes from the observationally determined solar visual magnitudes $V_{\odot}$ and $M_{V,\odot}$ using $BC_V = M_{Bol}-M_V=m_{Bol}-V$.

2- Assume an absolute bolometric magnitude ($M_{Bol,\odot}$) for the Sun, and then calculate its bolometric correction ($BC_{V,\odot}$) from its absolute visual magnitude ($M_{V,\odot}$).

After choosing one of the two methods above, equation (2) could be used for calculating the $M_{Bol}$ of a star if its $L$ is known. Consequently, the star's $BC_V$ could be calculated from its visual magnitudes if its distance and interstellar extinction are available. After collecting a sufficient number of $BC_V$ from all types of stars, a $BC_V-T_{eff}$ relation could be established; this is a useful relation for estimating the $BC_V$ of single stars which could be used for calculating the $L$ of other stars from their observed $V$ magnitudes if their distances and extinctions are known. Most $BC_V-T_{eff}$ relations are in tabular format; only \citet{Flower96} and \citet{Eker20} gave it as a function of $\log T_{eff}$ in a polynomial format. 

Although the word ``assume'' in the beginning of both methods implies absolute arbitrariness, as soon as a decision is made ($BC_{V,\odot}$ value in the case 1, or $M_{Bol,\odot}$ value for the case 2), the problem reduces to limited discrete arbitrariness for users who access tabulated tables of $BC_V$ from several producers. For a producer, however, as long as the assumed value is treated as a unique quantity, all the other parameters ($BC_V$ and $M_{Bol}$) of other stars could be defined uniquely, fulfilling the fact that stars must have unique $BC_V$ and $M_{Bol}$.  

Historically, both methods are equally popular. \citet{Flower96} and \citet{Cox00} could be considered among the latest who assumed $BC_{V,\odot}=-0.08$ mag first, then calculated $M_{Bol,\odot}=4.74$ mag from $M_{V,\odot}=4.82$ mag, itself calculated from $V_{\odot}=-26.75$ mag, which is the primary observational quantity. \citet{Bessell98} gives a table (Table A4) where various authors \citep{Allen76, Durrant81, Schmidt82} who assumed $M_{Bol,\odot}$ first, then $BC_{V,\odot}$ calculated from $M_{V,\odot}$ which is itself obtained from $V_{\odot}$ together with corresponding solar fluxes and luminosities. \citet{Torres10} and \citet{Casagrande18} are the most recent examples where stellar $BC_V$ was obtained by adopting $M_{Bol,\odot}=4.75$ mag, while \citet{Eker20} preferred to adopt $M_{Bol,\odot}=4.74$ mag since it is a nominal value suggested by IAU 2015 General Assembly.

Both methods are quite well acceptable and eligible to produce a consistent unique $BC - T_{eff}$ curve. In addition to the discrete arbitrariness (users have several choices), what is more problematic is that users must be alerted when choosing {\it BC} corrections from literature. They must first search what values of $M_{Bol,\odot}$ and $L_{\odot}$ were adopted at the very beginning because the same adopted values are needed when calculating the $L$ of other stars via equation (2). Even a slight change on any one or both of those adopted values will be reflected in as systematic errors on the computed $L$. The risk of error has already been point out by \citet{Torres10} ``When making use of tabulations of empirical bolometric corrections for stars ($BC_V$), a commonly overlooked fact is that while the zero point is arbitrary, the bolometric magnitude of the Sun ($M_{Bol,\odot}$) that is used in combination with such tables cannot be chosen arbitrarily''.

In order to avoid such problems and also remove discrete arbitrariness, consensus on a method and its assumed parameter is needed. Such a consensus has recently been achieved by IAU 2015 General Assembly Resolution B2.    

\subsubsection{IAU 2015 GAR B2}
It is not possible to say that the IAU has not issued a formal resolution on the matter of $BC_V$ zero points because two of its commissions did agree only on a zero point of the bolometric magnitude scale by adopting a value for $M_{Bol,\odot}=4.7$5 mag at the Kyoto meeting of 1997 \citep[][pp. 141 and 181]{Andersen99}. On one occasion, the bolometric magnitude scale was set by defining a star with $M_{Bol}=0.00$ mag for an absolute radiative luminosity $L=3.055\times10^{28}$ W \citep[see also][]{Cayrel02}. However, as \citet{Casagrande18} commented, since $L_{\odot}$ is a measured quantity while $M_{Bol,\odot}$ is an arbitrary zero point, any value of $M_{Bol,\odot}$ is legitimate on the condition that once chosen, all bolometric corrections are scaled accordingly. That is, no extra consensus is needed to pin down the value of the zero point constant of the $BC_V$ scale. Indeed, fixing the zero point of the bolometric magnitude scale has the consequence of fixing the zero points of the all {\it BC} scales of all other bands, not only $BC_V$.    

Arbitrariness of various {\it BC} scales, apparently, was an issue discussed first by IAU commissions. Nevertheless, the decisions on the subject were finalized by IAU 2015 General Assembly Resolution B2 (after IAU 2015 GAR B2). The zero point constant of bolometric magnitudes was set according to: 
\begin{equation}
M_{Bol}=-2.5\log (L) + C_{Bol}.
\end{equation}
with the clear statement ``{\it a radiation source with absolute bolometric magnitude $M_{Bol}=0$ has a radiative luminosity of exactly $L_0=3.0128\times10^{28}$ W and the absolute bolometric magnitude $M_{Bol}$ for a source of luminosity $L$ (in W) is:  
\begin{equation}
M_{Bol}=-2.5\log (L/L_0) = -2.5 \log L + 71.197~425~...~.
\end{equation}
The zero point was selected so that the nominal solar luminosity corresponds closely to absolute bolometric magnitude $M_{Bol,\odot}=4.74$ mag, the value most commonly adopted in the recent literature} {\it \citep*[e.g.][]{Bessell98, Cox00, Torres10}}''\footnote{https://www.iau.org/static/resolutions/IAU2015 English.pdf} together with a reminder: the notations of $M_{Bol}$ and $m_{Bol}$ were adopted by Commission 3 (Notations) at the 6th IAU General Assembly in Stockholm in 1938\footnote{https://www.iau.org/static/resolutions/IAU1938 French.pdf}. It is clearly declared that $M_{Bol}$ and $m_{Bol}$ refer specifically to absolute and apparent bolometric magnitudes respectively.  Thus, the zero point constant of the bolometric magnitude scale is $C_{Bol}= 71.197~425~ ...$ if $L$ is in SI units, and $C_{Bol}=88.697~425~...$ if $L$ is in cgs units. 

Equation (3) is more fundamental than equation (2) because if equation (3) is written for the Sun:

\begin{equation}
M_{Bol,\odot}=-2.5\log L_{\odot} + C_{Bol}.
\end{equation}
Subtracting this equation side by side from equation (3), the equation (2) is obtained. Contrary to equation (2), which is the source of arguments favouring arbitrariness, the more fundamental relation equation (3) leaves no space for any kind of arbitrariness if the value of $C_{Bol}$ is known. In equation (2), the term $C_{Bol}$ cancels out during the subtraction. From equation (5) it can be understood why the Sun was chosen for setting the zero point constant ($C_{Bol}$) of the bolometric magnitudes. This is because the most accurate stellar luminosity ever determined is solar luminosity.   

The best estimate of solar luminosity is $L_{\odot}=4\pi({\rm 1~ au})^2S_{\odot}=3.8275(\pm0.0014)\times10^{26}$ W, which is a value computed from the recently estimated best value of the solar constant $S_{\odot}=1361(\pm1)$ Wm$^{-2}$, obtained from the total solar irradiance (TSI) data of 35 years of space-born observations during the last three solar cycles \citep{Kopp14}. The adopted values of $C_{Bol}$ and $L_{\odot}$ indicate the nominal value of $M_{Bol,\odot}=4.739~996~ ...$ mag, which is indeed very close to $M_{Bol,\odot}=4.74$ mag as specifically pointed out by IAU 20015 GAR B2.     

\subsubsection{Other Types of Arbitrariness?}

The discrete arbitrariness of the zero point of the $BC_V$ scale is removed after assigning a definite value to the zero point constant for bolometric magnitudes. If $M_{Bol}$ and $L$ have one-to-one correspondence, both could be considered as unique properties of a star. Since visual absolute magnitude $M_V$ is also a unique property of this star, the basic definition, $BC_V=M_{Bol}-M_V$ is there to produce unique $BC_V$. This fact, however, became noticeable to us after IAU 2015 GAR B2; that is, the arbitrariness of type 0 attributed to the zero point of the $BC_V$ scale had been removed.

The tabulated tables of $BC_V$ actually define a relationship between $BC_V$ coefficients and other stellar parameters, even though there is no definite form for a function. The primary parameter influencing $BC_V$ is effective temperature. \citet{Flower96} and \citet{Eker20} give the $BC_V-T_{eff}$ relation as polynomials, using the logarithm of $T_{eff}$ as a single variable. Therefore, it can be said that there is no arbitrariness of Type 1 associated with $BC_V$ values as well as $BC_V-T_{eff}$ curves. Moreover, there should not be an arbitrariness problem with Type 2 either because the least squares method guarantees a unique fitting curve ($BC_V-T_{eff}$ function) with arbitrary coefficients determined, including internal random errors. The parameter space of the $BC_V-T_{eff}$ function is also unique; hence there is no space for a Type 3 problem because for a given $T_{eff}$ there is only one value of $BC_V$. Thus,  the arbitrariness problems of Type 0, Type 1, Type 2 and Type 3 do not exist for $BC_V$ and $BC_V-T_{eff}$ relations in reality. 

\subsection{Apparent Magnitudes and Zero Point Constants}
A comprehensive description of the magnitude system is given concisely by \citet{Cox00} and various textbooks. There are various filters to isolate certain spectral features or wavelength ranges. The fluxes received through these filters are transformed to magnitude values. For monochromatic radiation, a flux density per unit wavelength of $3.631\times10^{-9}$ erg s$^{-1}$cm$^{-2}$ \AA$^{-1}$ defined as m = 0.0 \citep{Casagrande14}, equivalent to -21.10 mag, which is known as the zero point constant of ST magnitudes. If the monochromatic magnitudes use fluxes of per unit frequency, the system is called the AB system. For AB magnitudes, a flux of $3.631\times10^{-20}$ erg s$^{-1}$cm$^{-2}$ Hz$^{-1}$ is set up for $m = 0.0$ magnitude, where the zero point constant becomes -48.60 mag \citep{Casagrande14}. \citet{Bessell98} provide them to three digits accuracy: $C_{\lambda}=-21.100$, and $C_{\nu}=-48.598$. The fluxes which make $m=0.0$ were chosen so that, for convenience, $\alpha$ Lyr (Vega) has very similar magnitudes in all systems: ST, AB and VEGA \citep{Casagrande14}. 

In reality, the magnitudes of various photometric systems are heterochromatic. Using $\alpha$ Lyr (Vega) as the primary calibrating star, the VEGA system is the most well known and deliberated for heterochromatic measurements. Among these systems, the most famous is the Johnson-Cousins system. Although the zero points are often determined observationally from a network of standard stars, it is formally just a single object. A hypothetical star of the spectral type A0V with magnitude $V=0.0$ mag on the Johnson system is given in Table 16.6 \citep{Cox00}, where {\it UBVRI} bands and monochromatic fluxes at effective wavelengths are presented. Note that Vega is used as a calibrating star but its apparent magnitude is not exactly zero. $V=0.03$ mag has been measured by \citet{Johnsonetal66} and \citet{Bessell98}. The standard Johnson value of $V=0.03$ mag is cited by \citet{Bohlin14}. The same value ($V=0.03$ mag) was adopted by \citet{Cox00}, \citet{Girardi12}, \citet{Bessell12} and \citet{Casagrande14}. For the other bands, Vega is found to be just slightly positive ($\sim0.02$ mag) at most bands \citep{Rieke08}.

For the heterochromatic bands of various photometric systems, the zero point constant $C_{\xi}$ needs to be derived for each bandpass $\xi$  using a star of known absolute flux, usually Vega \citep{Casagrande14}, or Sirius and Vega \citep{Bohlin14}. Although $C_{\xi}$ are usually not given for the photometric systems in the literature, where only monochromatic fluxes at effective wavelengths of the filters are listed, all zero point constants are well defined quantities \citep{Bessell98, Cox00, Girardi12, Casagrande14}.

The advantage of the magnitude system is that it works even if $C_{\xi}$ are unknown. Since zero point constants cancels out, only ratio of fluxes can be determined by subtracting apparent magnitudes; then knowing the absolute flux of one star is sufficient to calculate the absolute flux of the other star from their magnitude differences. On the other hand, unique (unchanging) $C_{\xi}$ for each band is there to define the intrinsic colour indexes (if there are no interstellar and atmospheric effects). Therefore, photometric bands and their special zero points (definite fluxes unique to each band) are natural for the astronomical photometry to work. 

Apparent visual magnitude of a star could be expressed as: 
\begin{equation}
V=-2.5\log f_{V}+C_{V} = -2.5 \log \int_0^\infty \!S_{\lambda}(V)f_{\lambda}\rm d\lambda + C_{V},
\end{equation}
where $f_V=\int_0^\infty \!S_{\lambda}(V)f_{\lambda}\rm d\lambda$
is the $V$ flux reaching to the telescope (no interstellar and atmospheric effects), $S_{\lambda}(V)$ is the $V$ band transparency function, $f_{\lambda}$ is the monochromatic flux coming from the star and $C_V$ is the zero point constant for the $V$ band. Similarly, it can be adopted for bolometric apparent magnitudes as: 
\begin{equation}
m_{Bol}=-2.5\log f_{Bol}+C_{Bol} = -2.5 \log \int_0^\infty \!f_{\lambda}\rm d\lambda + C_{Bol},
\end{equation}
where $f_{Bol}=\int_0^\infty \!f_{\lambda}\rm d\lambda$ is the bolometric flux reaching to the telescope (again, no interstellar and atmospheric effects) and $C_{Bol}$ is the zero point constant for the bolometric apparent magnitudes. Using the nominal solar values, $m_{Bol,\odot}= 26.832$ mag \citep{Cox00}, and $f_{Bol,\odot}=1361$ Wm$^{-2}$ (IAU 2015 GAR B2), one can calculate the value of the zero point constant
$C_{Bol}=-18.997~351~...$ mag for the apparent bolometric magnitudes. Consequently, $f_{Bol}=2.518~022~... 10^{-8}$ Wm$^{-2}$ could be calculated using the same $C_{Bol} = -18.997~351~...$ mag for making $m_{Bol}=0$.

Bolometric apparent magnitude appears as if there is brightness at another band, in addition to the other bands of the Johnson photometry. This hypothetical brightness, however, represents a band of radiation so broad that it covers all wavelengths; thus, it represents the total radiation of a star. As is calculated   for obtaining the intrinsic colour of an unreddened star, subtracting side by side equation (6) from equation (7), according to its basic definition, $BC_V$ becomes:   

\begin{eqnarray}
BC_V = m_{Bol}-V = 2.5\log \frac{f_V}{f_{Bol}}+(C_{Bol}-C_V)\\ \nonumber
= 2.5\log \left(\frac{\int_0^\infty \!S_{\lambda}(V)f_{\lambda}\rm d\lambda}{\int_0^\infty \!f_{\lambda}\rm d\lambda}\right)+C_2.
\end{eqnarray}

Here is the answer to the question of why there is an arbitrary constant in equation (1), despite the definite integrals not taking any constant. $C_2$ is certainly not a constant of integration, even though it appears as if it is one. It is a natural result of absolute photometry. Also, it is not correct to call it ``arbitrary'' anymore, because $C_2= C_{Bol}-C_V$, where $C_{Bol}$ and $C_V$ are the definite zero point constants of the bolometric and visual magnitudes according to equations (7) and (6).   

\subsection{Absolute Magnitudes and Zero Point Constants}
Absolute magnitudes are also expressed by the same notation as apparent magnitudes. Thus, the absolute magnitude of a star at the photometric band $\xi$ is expressed as:

\begin{equation}
M_{\xi}=-2.5\log F_{\xi} + C_{\xi}.
\end{equation}
and the bolometric absolute magnitudes as:
\begin{equation}
M_{Bol}=-2.5\log F_{Bol} + C_{Bol}.
\end{equation}
then, $M$ and $F$ stand for absolute magnitudes and fluxes if a star is moved to a distance of 10 pc. Now the question is: What happens to the zero point constants $C_{\xi}$ and $C_{Bol}$? Are they the same constants used for apparent magnitudes? 

It is logical to assume the zero point constants of absolute magnitudes are the same as the zero point constants of apparent magnitudes because the zero point flux for a photometric band, which marks zero magnitude on the magnitude scale, is independent of the distance. But still this can be tested by taking the $C_{Bol}$ of the apparent bolometric magnitudes given above, and using $M_{Bol,\odot}=4.74$ mag, 
\begin{equation}
4.74=-2.5\log F_{Bol,\odot} - 18.997~351~...~,
\end{equation}
then the bolometric flux of the Sun, if it is at a distance of 10 pc $F_{Bol,\odot}=3.199~334~... \times 10^{-10}$ W m$^{-2}$ is obtained. If this flux is multiplied by the surface area of a sphere with a radius of 10 pc, it gives $L_{\odot}=3.828\times10^{26}$ W. This is the nominal solar luminosity announced by IAU GAR B2, which was calculated from the solar constant $S_{\odot} =1361(\pm1)$ Wm$^{-2}$. Therefore, we can claim confidently that the zero point constants of apparent (equation 7) and absolute bolometric magnitudes (equation 10) are the same $C_{Bol} {\rm (fluxes)}=-18.997~351~...$~.  Moreover, the same conclusion could be obtained by $BC_{\xi}=M_{Bol}-M_{\xi}=m_{Bol}-m_{\xi}$. 

Using the part
\begin{equation}
M_{Bol}-M_{\xi}=m_{Bol}-m_{\xi}
\end{equation}
and moving $m_{\xi}$ to the left, then moving $M_{Bol}$ to the right algebraically, the equation becomes: 
\begin{equation}
m_{\xi}-M_{\xi}=m_{Bol}-M_{Bol}.
\end{equation}
Both equations (12 and 13) are valid. The former is the basic definition of $BC_{\xi}$ while the latter is the distance modulus, which is equal to ($5\log d-5$), where $d$ is in parsecs. Since the wavelength dependent zero point constants are cancelled between the equal signs separating various photometric bands, equation (13) tells us that the distance modulus is independent of wavelengths. This could happen only if the zero point constant for apparent and absolute magnitudes have the same value (Table 1). While the distance modulus is independent of wavelengths, conversely, the bolometric corrections defined by equation (12) are dependent on wavelengths (or bands). Thus, bolometric correction is a unique feature of a star, as with its intrinsic colours. 

\begin{table*}
  \centering
  \caption{Zero point constants for calculating stellar luminosities or fluxes.}
    \begin{tabular}{ccccccc}
    \hline
    $C_{Bol}$ for $L$ & $L_0$ & Unit  &  $C_{Bol}$ for $f_{Bol}$ & $C_{Bol}$ for $F_{Bol}$ & $f_0$    & Unit \\
    \hline
    71.197~425~... & 3.0128E+28 & W            & -18.997~351~... & -18.997~351~... & 2.5180E-08 & W m$^{-2}$ \\
    88.697~425~... & 3.0128E+35 & erg s$^{-1}$ & -11.497~351~... & -11.497~351~... & 2.5180E-05 & erg s$^{-1}$ cm$^{-2}$ \\
    \hline
    \end{tabular}%
  \label{tab:addlabel}%
\end{table*}%

\subsection{From a Flux Ratio to a Luminosity Ratio}
Taking one of the $M_{\xi}$ (equation 9), let it be $M_V$, and by subtracting it side by side from $M_{Bol}$ (equation 10), it can be written:
\begin{equation}
M_{Bol}- M_{V} = 2.5\log \frac{F_V}{F_{Bol}}+(C_{Bol}-C_V).  
\end{equation}
where, $F_V/F_{Bol}$ is the ratio of visual flux to bolometric flux reaching to Earth (no atmospheric and interstellar effects) if the star is moved to a distance of $d=10$ pc. According to the basic definition of $BC_V$, this equation and equation (8) are equivalent. Hence, it can be written: 
\begin{equation}
\frac {F_V}{F_{Bol}} = \frac {f_V}{f_{Bol}}.  
\end{equation}
which means that the ratio of visual flux to bolometric flux for a star is independent of its distances. Therefore, some more equal signs could be added:
\begin{equation}
\frac {F_V}{F_{Bol}} = \frac {f_V}{f_{Bol}}= \frac {f_V(S)}{f_{Bol}(S)}= \frac {4\pi R^2f_V(S)}{4\pi R^2f_{Bol}(S)} = \frac {L_V}{L}=\frac{f_V}{\sigma T^4}.  
\end{equation}
The third term is the flux ratio on the surface of the star. Multiplying the surface fluxes (visual and bolometric) by $4\pi R^2$ (surface area of a star), then its equivalent, the luminosity ratio ($L_V/L$), could be added. Finally, by replacing the surface bolometric flux by $\sigma T^4$, one can see the primary parameter (effective temperature) to determine the {\it BC} values of stars according to the rightmost ratio. After replacing $F_V/F_{Bol}$ in equation ( 14) by $L_V/L$, one can deduce:  
\begin{equation}
M_{Bol}=-2.5\log L+C_{Bol},
\end{equation}
and,
\\
\begin{equation}
M_{V}=-2.5\log L_V+C_{V}.
\end{equation}

We must be careful here; the $C_{Bol}$ providing $M_{Bol}$ of stars from their $L$ is different ($C_{Bol} =71.197~425~...$ mag) from the $C_{Bol}$ providing the same $M_{Bol}$ from its bolometric flux if it is at 10 pc [$C_{Bol}{\rm (fluxes)}= -18.997~351~...$ mag]. Similar case must be true for the apparent magnitudes. Subtracting equation (18) from equation (17), the following could also be written for $BC_V$
\begin{equation}
BC_{V}= 2.5\log \frac{L_V}{L}+C_2.  
\end{equation}
where $C_2=C_{Bol}-C_V$. Here, one must be aware of the fact that the zero point constants proper to the luminosities (equations 17 and 18), which are different from the zero point constants for fluxes (equations 6, 7 and 9, 10).  

\subsection{Analytical Evidence for Limited Arbitrariness and Positive {\it BC}}
According to equations (1), (8) and (19), the arbitrariness attributed to the zero point constant ($C_2$) of the $BC_V$ scale cannot be called absolute arbitrariness but it could be called limited arbitrariness. In other words; no value of $C_2$ from minus infinity to plus infinity (absolute arbitrariness) satisfies these equations. This is because the visual to bolometric flux ratio, or ratio of luminosities $L_V$ to $L$, is a unitless number [$(0<f_V/f_{Bol}<1)$ or $(0<L_V/L<1)$] between zero and one. Logarithms of numbers between zero and one are all negative. Only a positive $C_2$ would satisfy equations (1), (8) and (19) if the $BC_V$ of a star is zero, which occurs at $M_{Bol}=M_V$. Therefore, from equations (1) and (14): 
\begin{equation}
C_2(V)=-2.5\left|\log \frac{f_V}{f_{Bol}}\right|_{BC_{V}=0}. 
\end{equation}
This equation could easily be adapted for $C_2(\xi)$ by replacing $f_V$ by $f_{\xi}$. Thus, each band has to be evaluated at the point $BC_{\xi}=0$, which occurs at $M_{Bol}=M_{\xi}$. Since the flux of each band, not only the visual flux, is less than the bolometric flux, the right hand side of equation (20) is a positive number. The ratio in the logarithmic term has no chance to be either zero or one, therefore, $C_2(\xi)$ is strictly a number greater than zero. 

The $C_2(\xi)$ is still arbitrary depending upon which star is chosen for $BC=0$ mag, but the arbitrariness implied by equations (1), (8) and (19) is not an absolute arbitrariness, which would also include zero and negative numbers. Since only positive numbers are possible for $C_2(\xi)$, the type of the arbitrariness is limited.   

After knowing the fact that $C_2(\xi)>0$, it becomes easy to investigate whether a positive {\it BC} exists or not. From equation (20) the following could be deduced: if the absolute values of the logarithmic terms in equations (1), (8), and (19) 
\begin{equation}
\left|2.5\log \frac{f_{\xi}}{f_{Bol}}\right|<C_2(\xi). 
\end{equation}
is less than the value of the zero point constant $C_2 (\xi)$, the $BC_{\xi}$ values are positive, if
\begin{equation}
\left|2.5\log \frac{f_{\xi}}{f_{Bol}}\right|>C_2(\xi). 
\end{equation}
the $BC_{\xi}$ values are negative. It can therefore be concluded here that a limited number of positive {\it BC} values is inevitable.

A limited number of positive {\it BC} may also be deduced from following: 

\begin{equation}
L_V = L \times 10^{\frac{BC_V-C_2}{2.5}}. 
\end{equation}
which is another form of equation (19). All of the $BC_V$ values smaller than $C_2$, including  zero and limited range of positive numbers ($0\leq BC_V<C_2$), are valid to produce a $L_V$ which is less than $L$. $BC_V=C_2$ would mean $L_V=L$ and $BC_V>C_2$ would produce $L_V>L$, which are invalid cases because $L_V\geq L$ is unphysical.

\section{Discussion}
\subsection{Positive Bolometric Corrections; Problematic or not Problematic?} 
There should not be any other option in between; a positive {\it BC} is either a conceptual problem, or not. No one should say ``whether the {\it BC} values are positive or negative is subject to the caprice of astronomers'' and then ask the question ``Does anybody really care whether {\it BCs} ``have to be'' positive or not?''

Positive science does not operate on the caprice of scientists and care must be given not to stray from scientific consistency. A positive {\it BC} was already said to contradict the fact that ``the bolometric magnitude of a star ought to be brighter than its $V$ magnitude'' \citep{Torres10}. Moreover, to emphasize the importance for {\it BC} numbers to be negative, \citet{Cox00} listed the smallest {\it BC} of stars (main-sequence, giants and supergiants) as ``-0.00'' for F2 supergiants in the astronomers' handbook Allen's Astrophysical Quantities, 4th edition, on page 389. The negative sign appearing as ``-0.00'' indicates that {\it BCs} are not even neutral (zero) but strictly negative numbers. ``-0.00'' implies that the {\it $BC_V$} of F2 supergiants could contain a numeral other than zero, with at least two zeros after the decimal point.

A positive {\it BC} is also related to the arbitrariness problem attributed to the zero point of the {\it BC} scale. Arbitrariness, especially absolute arbitrariness, provides {\it BC} producers the freedom to make a systematic shift of $BC_V-T_{eff}$ diagram or on the tabulated {\it BC} numbers to make them all negative in order to avoid a dilemma. Two schools of thought have emerged. One group of astronomers intentionally apply a systematic shift to avoid even a small single positive {\it BC} (because this solves the problem of contradicting the fact that ``the bolometric magnitude of a star ought to be brighter than its $V$ magnitude''), while the other group of astronomers keep positive {\it BC} in their list and serve them to their users. The group keeping positive {\it BC} appears problematic if the arbitrariness problem is not solved; the group could be called agnostic or uncaring about the problems of positive {\it BC}. If indeed there is a solution to the arbitrariness problems, users of $BC_{V}-T_{eff}$ relations or tabulated tables of {\it BC} do have a right to hear what the solution is. If there is a solution, then the former group becomes inconsistent. There is no way in between, either one of the groups must give up its practice for consistent science, even if their practice is several decades or centuries old. 

\subsection{Theoretical and Observational Support for a Limited Number of Positive {\it BC}}

The analytical evidence for the limited arbitrariness and positive {\it BC} presented in Section 2.5 could be considered as theoretical support for a limited number of positive {\it BC} values, which appear from time to time in the literature. As discussed in Section 2.5, although a positive {\it BC} is inevitable analytically, there is no indication where (or at what temperature) the zero point of the {\it BC} scale for different photometrical bands would occur. This is the kind of problem where one should be able to see the existence of a limited number of positive {\it BC} values, but leave the astronomer with the freedom to define the zero point optionally, at best conventionally. The applicability and necessity of the decision by IAU 20015 GAR B2 is founded firmly in this freedom. In other words, the zero point of the {\it BC} scale suggested by IAU 20015 GAR B2 would be consistent with the analytical prediction if the zero point is not at the peak (or above the peak) of the $BC-T_{eff}$ curve. It has already been shown in Section 2.5 that unphysical results such as $L_V=L$ would occur if the zero point of the {\it BC} scale is placed right on the peak of the curve, which means $BC_V = C_2$. This case, however, is impossible according to equation (20). The unphysical result $L_V>L$ occurs if the zero point is placed above the peak, that is, if $BC_V>C_2$, where $C_2$ stands for the zero point constant for the $BC_V$ scale. Physically consistent results require $BC_V<C_2$, which includes both positive ($0<BC_V<C_2$)  and negative ($BC_V<0$) bolometric corrections.         

All empirical $BC_{V}$ vales computed from the observed stellar parameters \citep*{Code76, Johnson64, Johnson66, Flower77, Flower96, Bessell98, Sung13, Casagrande18, Eker20} could be considered as observational support for the existence of a limited number of positive $BC_{V}$. The $V$ band empirical bolometric correction coefficients of main-sequence stars computed most recently by \citet{Eker20} are of special importance not only because they were computed from the most updated and/or most accurate stellar parameters, but also because the nominal values $ M_{Bol,\odot}=4.74$ mag, and $L_{\odot}=3.828\times10^{26}$ W were used in computing the $BC_V$ values of 400 stars, as suggested by IAU 2015 General Assembly Resolution B2. 

According to the results of \citet{Eker20}, the peak of the $BC_V-T_{eff}$ curve occurs at $T_{eff} = 6897$ K, at about a spectral type $\sim $ F1V with a peak value $BC_{V, max} = 0.95$ mag and the two zero points ($BC_V=0$ mag) occur at effective temperatures $T_{eff, 1} = 5859$ K, and $T_{eff, 2} = 8226$ K corresponding to the spectral types $\sim$ G2V and $\sim$ A5V, respectively. Thus, according to the $BC_V-T_{eff}$ curve of \citet{Eker20}, main-sequence stars within the temperature limits $5859 < T_{eff}<8226$ K have positive $BC_V$ and the rest, that is, the earlier and later spectral types, have negative $BC_V$.     

Therefore, here we can go beyond the norm to say ``bolometric correction of a star is not always negative''. Considering the widespread effect of the well-known axiom ``the bolometric magnitude of a star ought to be brighter than its $V$ magnitude'' among practitioners for about a century, this conclusion is not just a new result but a paradigm change, which is supported not only by theoretical and observational evidence but also according to the convention announced by IAU 2015 GAR B2. 

\subsection{Comparing $M_{Bol}$ and $M_V$ rather than $L$ and $L_V$ is misleading}

Comparing the bolometric magnitude of a star to its $V$ magnitude, and then to say which one is brighter, could be misleading because the human senses are not absolute. Thus, the perceptions of the eye could be deceiving. Even though there is one-to-one correspondence between luminosity and corresponding brightness in magnitude units, as they are displayed by equations (3), (17) and (18), the same conclusion cannot be drawn by comparing total luminosity (whole spectrum) to one of its parts ($L_V$ formed by visual photons).

First of all, luminosities are physical but magnitudes are not physical quantities. Comparing physical quantities is definitely more meaningful and definite than comparing non-physical, which is most probably subjective rather than being objective. Then, magnitude is the measure of human perception of a luminosity, which was first suggested by Hipparchus of Nicaea (190-120 BC) and reformulated by Norman Robert Pogson in 1854 by defining a first magnitude star as a star that is 100 times brighter than a sixth magnitude star \citep{Cox08}, which is still used today. Thus, eye comparisons of magnitudes are meaningful only if they are conducted in the same wavelength range within the visible part of the electromagnetic spectrum; otherwise, meaningless to indicate which one of two stars is more luminous due to different eye sensitivity at different wavelengths.

Since the magnitude scale is in reverse order, a brighter magnitude is shown by a smaller number. Consequently, the phrase that ``bolometric magnitude of a star ought to be brighter than its $V$ magnitude'' indicates that the bolometric magnitude of a star has a smaller numerical value than its $V$ magnitude. Then, subtracting a bigger number ($M_V$ or $V$) from a smaller number ($M_{Bol}$ or $m_{Bol}$), one can deduce a negative number for the bolometric correction of a star according to the basic definition ($BC_V = M_{Bol}-M_V=m_{Bol}-V$), which is confirmed only for stars of the earliest and latest spectral types but not for main-sequence stars of spectral types $\sim$ (G2-A5). Drawing attention to comparing $M_{Bol}$ and $M_V$ rather than $L$ and $L_V$, the phrase ``bolometric magnitude of a star ought to be brighter than its $V$ magnitude'' could also be considered misleading because the conclusion drawn by this comparison is not firmly established, that is, bolometric corrections are not always negative according to the observational and theoretical evidences discussed above.

\subsection{Visualizing a Fictitious Problem: Arbitrariness}

Assuming the absolute bolometric magnitude of the Sun ($M_{Bol,\odot}$) is arbitrary \citep{ Casagrande18}, when computing the absolute bolometric magnitude ($M_{Bol}$) for a star according to equation (2) means that the computed $M_{Bol}$ of the star is also arbitrary. This way, there could be an infinite number of $M_{Bol}$ values computed from a single luminosity. This situation is visualized in Figure 1 by three preferred $M_{Bol,\odot}$ values using the luminosities of 400 stars which were selected by \citet{Eker20} for computing their $BC_V$ values, from which the $BC_V-T_{eff}$ relation for nearby main-sequence stars is predicted.

\begin{figure*}
\begin{center}
\includegraphics[scale=1,keepaspectratio]{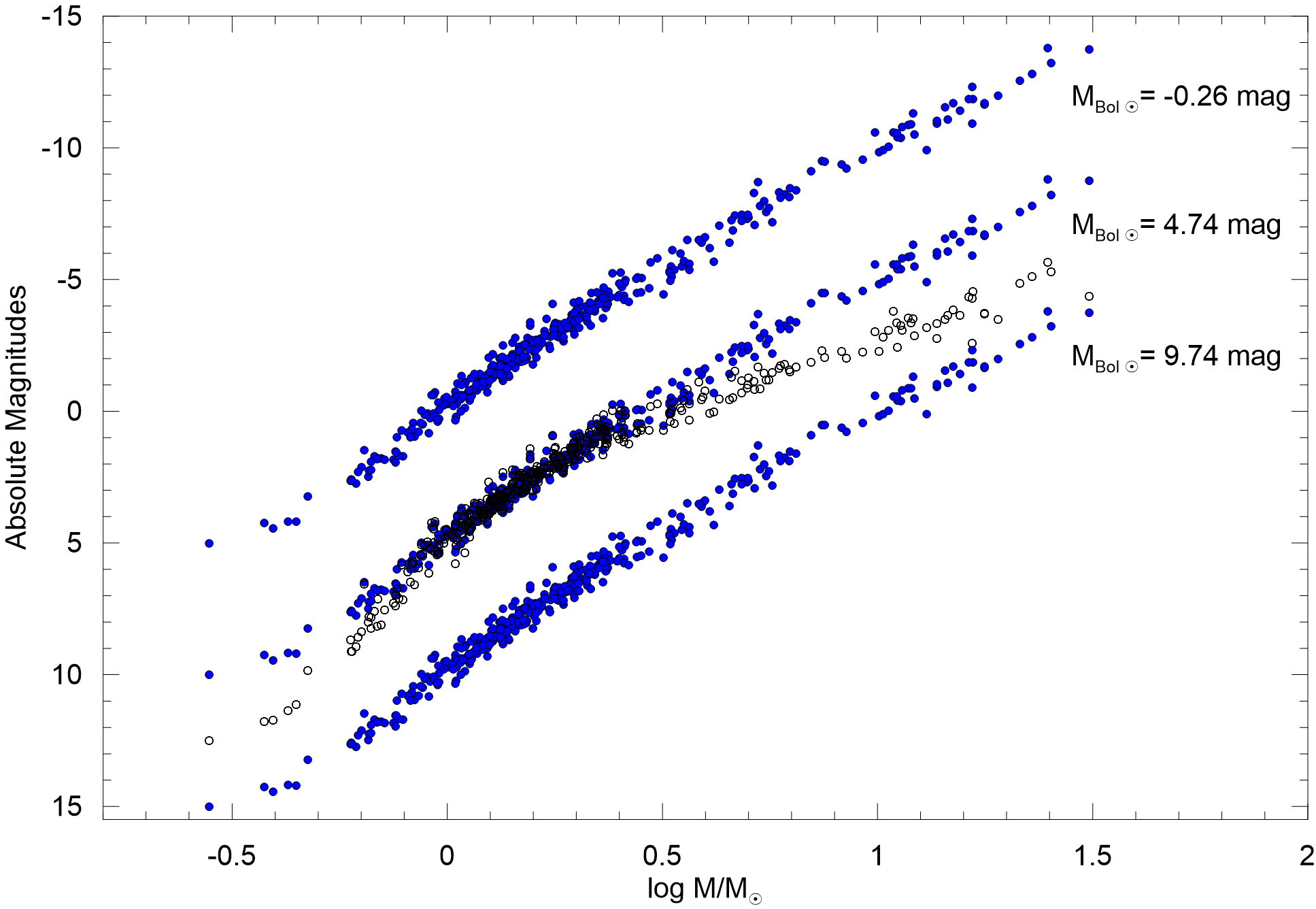}
\caption[] {Filled circles ($\bullet$) are $M_{Bol}$ of main-sequence stars \citep{Eker20} calculated according to equation (2) using $M_{Bol,\odot}=-0.26$ mag (upper), $M_{Bol,\odot}=4.74$ mag (middle), $M_{Bol,\odot}=9.74$ mag (lower). Empty circles (o) are absolute visual magnitudes $M_V$, (o) for the same stars.}
\end{center}
\end{figure*}

The filled and empty circles in the middle of Figure 1 show $M_{Bol}$ and $M_V$ values of the 400 main-sequence stars plotted as a function of stellar mass in solar units. Despite the stellar luminosities staying constant, because of the enforced arbitrariness \citep{Casagrande18}, the $M_{Bol}$ values are shifted up or down according to the adopted values of $M_{Bol,\odot}$ as indicated on the figure. Only two examples of all positive BC or all negative {\it BC} were indicated. That is, if one prefers to use $M_{Bol,\odot}=9.74$ mag, then all computed $BC_V$ values would be positive, while choosing $M_{Bol,\odot}=-0.26$ mag makes all $BC_V$ values negative (See Figure 2). 

\begin{figure*}
\begin{center}
\includegraphics[scale=1,keepaspectratio]{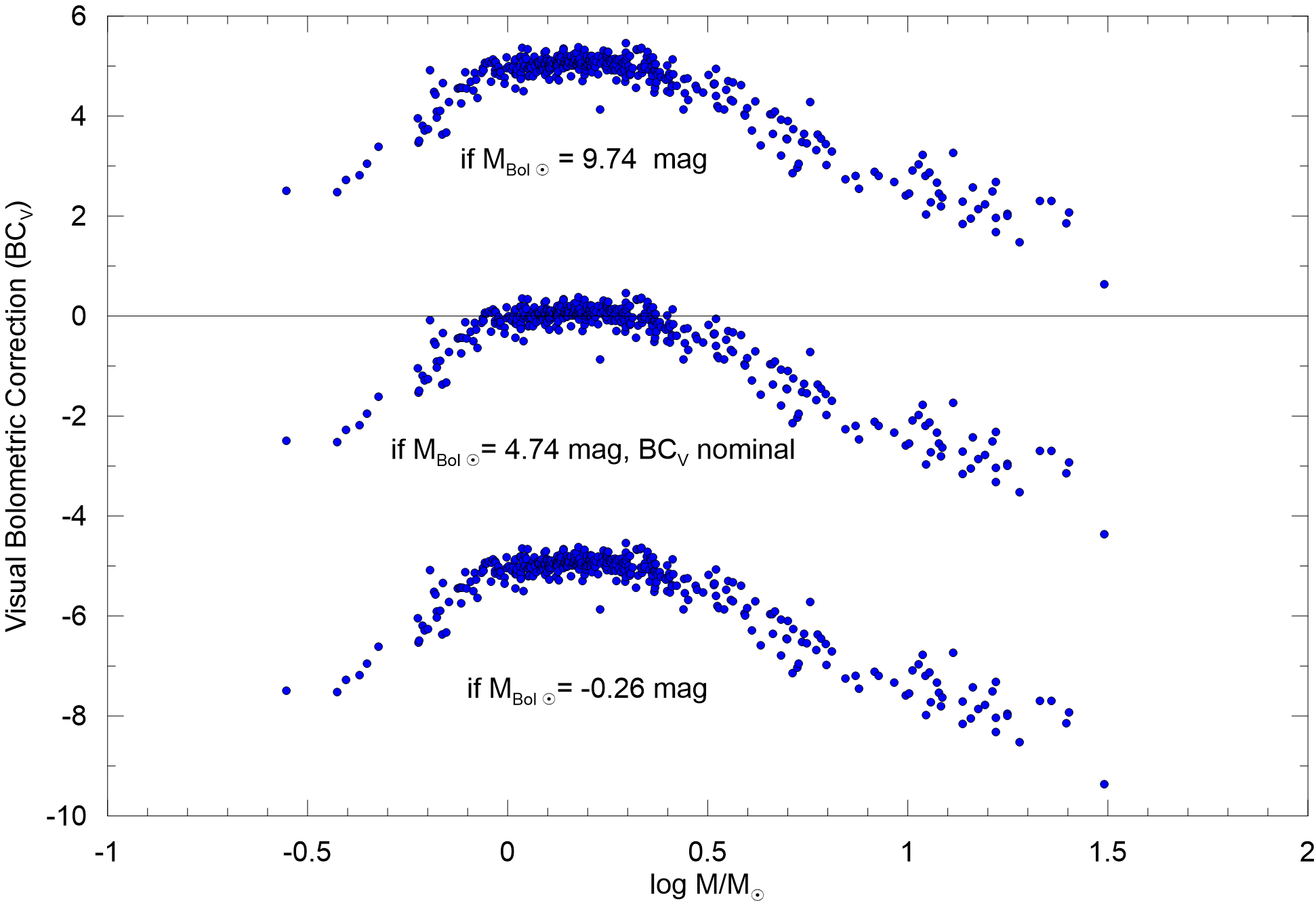}
\caption[] {All positive (upper $M_{Bol,\odot}=9.74$ mag), standard (middle $M_{Bol,\odot}=4.74$ mag) and all negative (below $M_{Bol,\odot}=-0.26$ mag) $BC_{V}$ coefficients of the main-sequence stars \citep{Eker20} as a function of $\log M/M_{\odot}$.}
\end{center}
\end{figure*}

If the absolute visual magnitudes ($M_V$) of stars in Figure 1 are subtracted from the corresponding $M_{Bol}$, then the corresponding $BC_V$ are obtained as a function of stellar mass (Figure 2). The same $BC_V$ data could be re-arranged according to stellar effective temperatures (Figure 3), which is the same $BC_V$ data from which \citet{Eker20} derived their $BC_V-T_{eff}$ relation. The similar appearance of $BC_V-M/M_{\odot}$ and $BC_V-T_{eff}$ is not surprising. 

\begin{figure*}
\begin{center}
\includegraphics[scale=1,keepaspectratio]{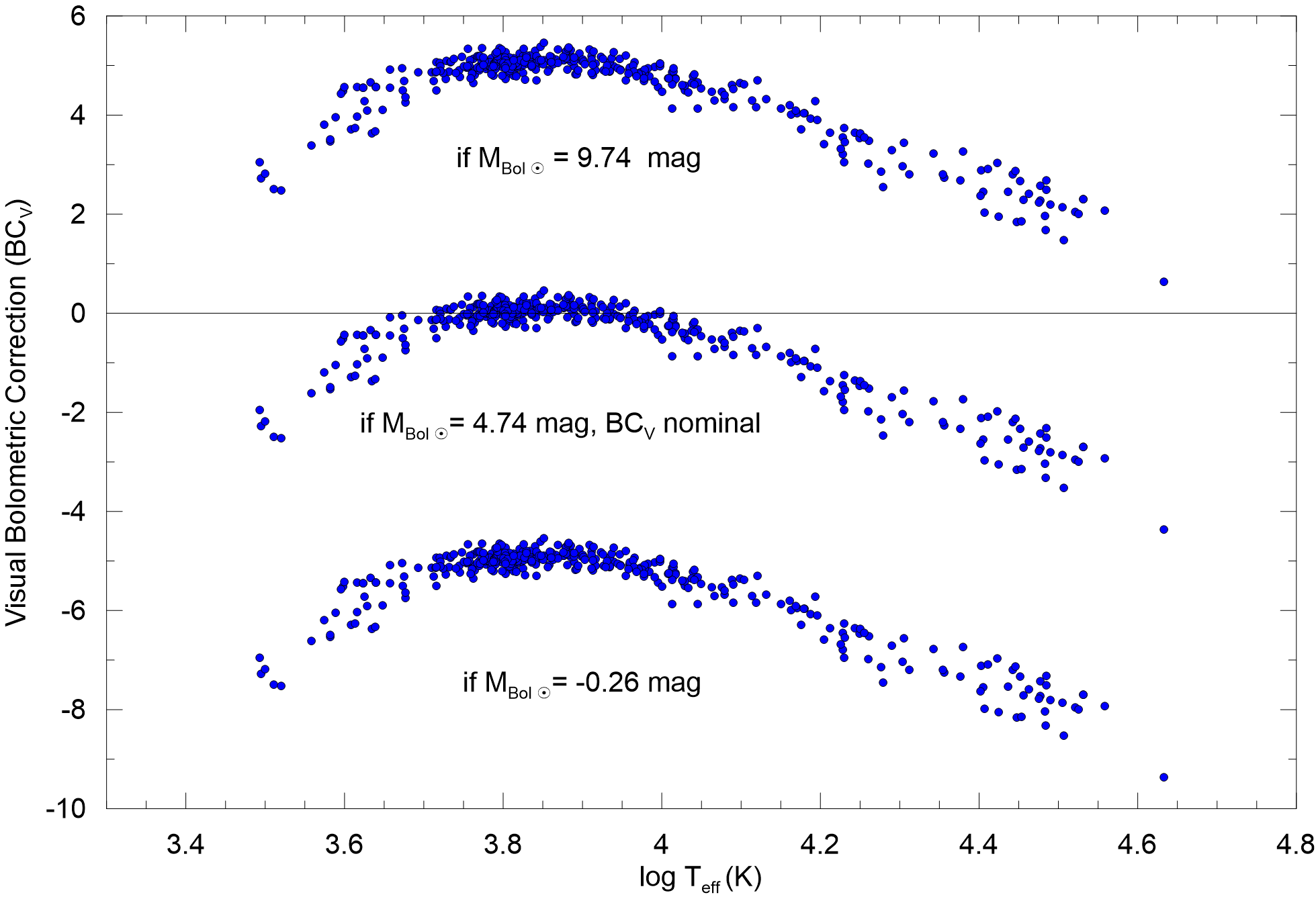}
\caption[] {All positive (upper $M_{Bol,\odot}=9.74$ mag), standard (middle $M_{Bol,\odot}=4.74$ mag) and all negative (below $M_{Bol,\odot}=-0.26$ mag) $BC_{V}$ coefficients of the main sequence stars \citep{Eker20} as a function of $\log T_{eff}$.}
\end{center}
\end{figure*}

It is by definition, that a star has a unique $M_{Bol}$, thus a unique $BC_V$, coming from its unique $L$ and $M_V$. Only after both absolute and limited discrete arbitrariness attributed to the zero point of {\it BC} scales is removed by IAU 2015 GAR B2 can such a consistency be achieved. The arbitrariness attributed to $M_{Bol, \odot}$, \citep{Casagrande18} violates this fact as displayed in Figures 1, 2, and 3, so the infinite number of $BC_V$ becomes inevitable. At this point, it can be said that the arbitrariness attributed to $M_{Bol, \odot}$ and the zero point of the {\it BC} scale is a fictitious problem caused by the method of computing according to the basic definition ($BC_V=M_{Bol}-M_V=m_{Bol}-V$), where the true value of $M_{Bol,\odot}$  was not known, thus assumed to be arbitrary before IAU 2015 GAR B2. 

It may also be noticed here that the positive and negative occurrences of standard $BC_V$ values are also marked on Figure 2 and Figure 3 where the values above the horizontal line are positive. The very limited number of positive occurrences is clearly apparent.

Stressing the uniqueness of a {\it BC} for a star at any band is no different than the uniqueness of its $M_{Bol}$, $M_V$, $L$, $R$ or $T_{eff}$; there should not be a function such as $BC-T_{eff}$ and/or $BC-M/M_{\odot}$, which could relate a {\it BC} value directly to a star's $T_{eff}$ or mass. Similar to the MLR, MRR and MTR functions of \citet{Eker18}, which are not the functions to provide the $L$, $R$ or $T_{eff}$ of a star given mass but to provide mean values of $L$, $R$ and $T_{eff}$ of main-sequence stars of given mass. According to stellar structure and evolution theories \cite{Clayton68} true $L$, $R$, and $T_{eff}$ are unique for a star which depend on its mass, chemical composition and age. In the same way of thinking, $BC-T_{eff}$ or $BC-M/M_{\odot}$ should provide typical (a kind of mean value) {\it BC} value for the typical effective temperature of main-sequence stars. Thus, the typical mass, typical effective temperatures and typical {\it BC} are related. The unique {\it BC} of a single star is provided only by the basic definition of {\it BC} ($BC_{\xi}=M_{Bol}-M_{\xi}=m_{Bol}-m_{\xi}$). Scatter of data on a $BC_V-T_{eff}$ relation are not really due only to observational errors, as in the case of SED represented by a Planck function, but also due to differences of mass, chemical compositions and age. Thus, the $BC_V-T_{eff}$ relation derived by \citet{Eker20} cannot be used to calculate the $BC_V$ of a star with a $T_{eff}$. It rather provides a mean value of $BC_V$ for main-sequence stars having a typical $T_{eff}$ value, which could be related to a typical mass according to the MTR of \citet{Eker18}, or could be derived by directly from the masses of stars, as shown in Figure 2 in this study. The tabulated {\it BC} tables are also not free from this problem. They  too provide a mean $BC_V$ for a given mean value of a stellar parameter.  

\subsection{Importance of Standard {\it BC}}

There are two ways to obtain a standard $BC_{\xi}$ for a star from its luminosity ($L$): 
\begin{itemize}
\item[1)] Obtain its $M_{Bol}$ from its $L$ according to equation (3) using $C_{Bol}=71.197~425~...$ if $L$ is in SI units, $C_{Bol}=88.697~425~ ...$ if $L$ is in cgs units.

\item[2)] Obtain its $M_{Bol}$ from its $L$ according to equation (2) using nominal solar bolometric absolute magnitude $M_{Bol,\odot}=4.74$ mag, and the nominal solar luminosity $L_{\odot}=3.8275\times 10^{26}$ W.
\end{itemize}

After having its unique (standard) $M_{Bol}$ from its unique L according to one of the two methods above, its absolute magnitude ($M_{\xi}$) in the band $\xi$ should be calculated properly from its de-reddened apparent magnitude ($m_{\xi}$) using its most accurate stellar parallax. Then, the difference $M_{Bol}-M_{\xi}$ could be called standard $BC_{\xi}$. Using any other $C_{Bol}$ in the method (1) and any other non-standard $M_{Bol,\odot}$ and/or $L_{\odot}$ values in the method (2) is sufficient to make it a non-standard ($ BC_{\xi}$). Finally, the standard {\it BC} of many stars, e.g. main-sequence stars, could be used to calculate the standard $BC_V-T_{eff}$ curve, as done by \citet{Eker20}.

Although both methods are equivalent, the method (1) is primarily advised since it is more straightforward, and does not demand additional knowledge other than $C_{Bol}$. Standard {\it BC} tables or standard $BC-T_{eff}$ curves are very important for standardizing stellar luminosities. It also saves time for users trying to find out from inspecting which value of $M_{Bol,\odot}$ were used at the beginning, and prevents them from any kind of making mistakes by using the improper $M_{Bol,\odot}$ and $L_{\odot}$ when re-computing unknown $L$. 

Considering the non-standard tabulated tables of $BC_V$, especially where all $BC_V$ are less than zero, our estimate is that there could be about $\sim$0.1 mag differences between standard and non-standard {\it BC} of a star. An approximate $\sim$0.1 mag difference, however, is equivalent to a $\sim$5\% systematic error in the distances (or trigonometric parallaxes) and about 10\% systematic error in the stellar luminosities. Such uncertainty was tolerable in the era before {\it Gaia}. However now, after {\it Gaia}, it is no longer tolerable. Among 206 binaries used for establishing the standard $BC_V-T_{eff}$ relation of nearby main-sequence stars by \citet{Eker20}, who relied on {\it Gaia} Data Release 2 distances \citep{Gaia18}, 120 systems ($\sim 60\%$) have 2\% or better accuracy in the trigonometric parallaxes.  

Reducing the uncertainties of stellar $L$ by standard {\it BC}  is not only important for internal structure and evolution theories, which demand not only the most accurate observational parameters (mass, radius, temperature, luminosity), but are also very important for galactic and extra-galactic astronomy, even cosmology and Hubble law; the luminosity functions of galaxies all require better stellar luminosities.

\section{Conclusions}
The zero point constant of any {\it BC} scale [$C_2(\xi)$] must be a positive number according to equations (1), (8), (14) and (19) due to the fact that the apparent flux ($0<f_{\xi}/f_{Bol}<1$) or absolute flux ($0<F_{\xi}/F_{Bol}<1$) or luminosity ($0<L_{\xi}/L<1$) of a star, which is limited by a filter, is less than the apparent bolometric, or absolute bolometric fluxes or  total luminosity of the star. A positive $C_2(\xi)$, however, require at least some of the $BC_{\xi}$ values to be positive. 

Consequently, shifting {\it BC} values to make them all less than zero is not necessary, despite any arbitrariness attributed to it. In this study, it became clear to us that the zero point constant $C_2$ in equation (1) is not a constant of integration, even though it looks like one. This is because definite integrals never take a constant of integration. In reality, it is a zero point constant imposed by definitions of bolometric and visual magnitudes appearing in the basic definition according to absolute photometry [see equations (8), (14)]. It is shown in this study that $C_2=C_{Bol}-C_V$. Since the zero points of existing photometric systems are all well-defined \citep{Casagrande14}, but only very recently has IAU 2015 GAR B2 defined the zero point of bolometric magnitudes, then earlier astronomers had no option but to assume it arbitrary. Only, after IAU 2015 GAR B2 is it understood to be well defined quantity because the definite values of $C_{Bol}$ and $C_V$ leave no space for it to be an arbitrary number, even though a definite value of $C_V$ cannot be found easily in the literature [{\it for each bandpass $C_{\xi}$ could be derived using a star of known absolute flux, usually Vega \citep{Casagrande14}, or Sirius and Vega \citep{Bohlin14}}] where usually monochromatic fluxes at effective wavelengths of filters for an apparent magnitude zero are given \citep{Bessell98, Cox00, Girardi12, Casagrande14}. 

A positive $C_2$ implies $C_{Bol}>C_V$. However one must be careful here. $C_{Bol}=71.197~425~...$ if $L$ is in SI units, $C_{Bol}=88.697~425~...$ if $L$ is in cgs units are for calculating $L$ from $M_{Bol}$, or vice versa, according to equation (3) or (17). The zero point constant for equations (7) and (10), where apparent and absolute bolometric fluxes reaching Earth (if no atmospheric and interstellar effects) are related to apparent and absolute magnitudes, is different. Perhaps, it is more proper to indicate it as $C_{Bol}({\rm fluxes})=-18.997~351~ ...$ mag (IAU 2015 GAR B2) corresponding to a flux of $2.518\times10^{-8}$ Wm$^{-2}$ to make the absolute and apparent magnitudes zero. Table 1 indicates which of the zero point constants are good for calculating  stellar luminosities or fluxes.     

Arbitrariness attributed to the zero point of the {\it BC} scale does not exist in reality. It is a fictitious arbitrariness caused by the method of computing or because the zero point of bolometric magnitudes was not yet fixed. Since $C_2 = C_{Bol}-C_V$ and if $C_{Bol}$ is not fixed, then $C_2$ become arbitrary. It is irrational to assume a star having more than one {\it BC} or $M_{Bol}$. Absolute arbitrariness attributed to the zero point of {\it BC} scale, however, contradicts this fact because it implies a star can have an infinite number of {\it BC} and $M_{Bol}$, which is not true. Comments such as ``... while the solar luminosity $L_{\odot}$ is a measured quantity, $M_{Bol,\odot}$ is an arbitrary zero-point and any value is equally legitimate on the condition that once chosen, all bolometric corrections are scaled accordingly'' are indeed incorrect because stars can have only one $M_{Bol}$, and the Sun cannot be the only exception to having an infinite number of absolute bolometric magnitudes.

\section*{Acknowledgements}
We are grateful to Dr Edwin Budding and anonymous referee whose discussions and comments was useful improving the revised manuscript. Authors thank to Mr. Graham H. Lee for careful proofreading of the text and correcting its English grammar and linguistics. This work was supported in part by the Scientific and Technological Research Council of Turkey (T\"UB\.ITAK) by the grant number 114R072. Thanks to the Akdeniz University BAP office is thanked for providing partial support of this research.

\section*{Data Availability}
The data underlying this article are available at https://dx.doi.org/10.1093/mnras/staa1659

\bsp	
\label{lastpage}
\end{document}